\documentstyle[amsfonts,preprint,aps]{revtex}

\begin{document}
\draft
\title{{\bf Chaotic Diffusion on Periodic Orbits:}\\
{\bf The Perturbed Arnol'd Cat Map}}
\author{{\bf Itzhack Dana and Vladislav E. Chernov
}}
\address{Minerva Center and Department of Physics, Bar-Ilan University, 
Ramat-Gan 52900, Israel}
\maketitle

\begin{abstract}
Chaotic diffusion on periodic orbits (POs) is studied for the perturbed
Arnol'd cat map on a cylinder, in a range of perturbation parameters
corresponding to an extended structural-stability regime of the system on
the torus. The diffusion coefficient is calculated using the following
PO formulas: (a) The curvature expansion of the Ruelle zeta function. (b)
The average of the PO winding-number squared, $w^{2}$, weighted by a
stability factor. (c) The uniform (nonweighted) average of $w^{2}$. The
results from formulas (a) and (b) agree very well with those obtained by
standard methods, for all the perturbation parameters considered. Formula
(c) gives reasonably accurate results for sufficiently small parameters
corresponding also to cases of a considerably nonuniform hyperbolicity.
This is due to {\em uniformity sum rules} satisfied by the PO Lyapunov
eigenvalues at {\em fixed} $w$. These sum rules follow from general
arguments and are supported by much numerical evidence.\newline
\end{abstract}


\begin{center}
{\bf I. INTRODUCTION}\\[0pt]
\end{center}

Understanding to what extent chaotic motion in Hamiltonian systems exhibits
random properties such as diffusion is a problem of both fundamental and
practical importance. The existence of deterministic chaotic diffusion has
been approximately established using a variety of approaches
\cite{bvc,cm,g6,cffk,mmp,df,j3,dmp,da,d1,d2,dt,a,acl,wnv,cgs,e,mr,c,as}. A
systematic approach is based on the hierarchy of periodic orbits (POs)
embedded in the chaotic region \cite{d1,d2,dt,a,acl,wnv,cgs,e,mr,c,as}. Let
us summarize the main ideas of this approach by considering, for
definiteness, the kicked-rotor maps on the cylinder
\begin{equation}
l_{m+1}=l_{m}+f(x_{m}),\ \ \ x_{m+1}=x_{m}+l_{m+1}\ \ 
\mathop{\rm mod}\ 1\ ,  \label{map}
\end{equation}
where $l$ is angular momentum, $x$ is angle, and the force function $f(x)$
satisfies $f(x+1)=f(x)$ and $f(-x)=-f(x)$. The diffusion coefficient for 
(\ref{map}) is formally defined by:
\begin{equation}
D=\lim_{m\rightarrow \infty }D_{{\cal E}}(m),\ \ \ \ D_{{\cal E}}(m)=
\frac{\langle (l_{m}-l_{0})^{2}\rangle _{{\cal E}}}{2m},  \label{D}
\end{equation}
where $\langle \ \rangle _{{\cal E}}$ denotes average over an ensemble 
${\cal E}=\{(x_{0},\ l_{0})\}$ of initial conditions in a chaotic region.
Well known problems in a reliable numerical calculation of $D$ are: (a) The
roundoff errors caused by the chaotic exponential instability. (b) The
ambiguity in the choice of ${\cal E}$ and the iteration time $m$. These
problems are systematically solved \cite{d1} by choosing ${\cal E}$ as the
ensemble ${\cal U}_{n}$ of all the primitive POs of period $n$ in the
chaotic region and $m=n$. For the map (\ref{map}), a PO $p\in {\cal U}_{n}$
is generally defined by initial conditions $(x_{0}^{(p)},\ l_{0}^{(p)})$
satisfying
\begin{equation}
l_{n}^{(p)}=l_{0}^{(p)}+w_{p},\ \ \ \ x_{n}^{(p)}=x_{0}^{(p)},  \label{PPO}
\end{equation}
where $n$ is the smallest integer for which (\ref{PPO}) holds with
integer $w_{p}$, the {\em winding number} of PO $p$. Since (\ref{map}) is
essentially periodic in $(x,\ l)$ with period $1$, $(x_{0}^{(p)},\
l_{0}^{(p)})$ can be restricted to a unit torus, $-0.5\leq x_{0}^{(p)},\
l_{0}^{(p)}<0.5$. The ensemble ${\cal U}_{n}$ may be viewed as an
invariant ``level $n$'' approximation of the chaotic region and the
diffusion rate {\em on} ${\cal U} _{n}$ is given by \cite{d1}:
\begin{equation}
D(n)=\frac{1}{2nN(n)}\sum_{p\in {\cal U}_{n}}w_{p}^{2}=\frac{1}{2nN(n)}
\sum_{w}N_{w}(n)w^{2},  \label{Dn}
\end{equation}
where $N(n)$ is the total number of POs in ${\cal U}_{n}$ and $N_{w}(n)$
is the number of POs in ${\cal U}_{n}$ with $w_{p}=w$. Similarly, one
can associate diffusion rates with subensembles of ${\cal U}_{n}$
having well-defined dynamical characteristics \cite{d1,d2}. For
uniformly hyperbolic systems, (\ref{Dn}) is expected to approximate well
the diffusion coefficient (\ref{D}) associated with generic ensembles of
aperiodic chaotic orbits. In fact, in the case of the cat and sawtooth
maps \cite{cm}, (\ref{Dn}) gives the exact value of $D$ for the cat maps
and approximates very well $D$ for the sawtooth maps \cite{d1}.\newline

The diffusion coefficient for generic chaotic ensembles in hyperbolic
systems is given by the exact PO formula \cite{cgs,c,g}:
\begin{equation}
D=-\frac{1}{2}\left. \frac{\partial ^{2}\zeta ^{-1}(\beta ,\ s)/\partial
\beta ^{2}}{\partial \zeta ^{-1}(\beta ,\ s)/\partial s}\right| _{\beta
=s=0}.  \label{De}
\end{equation}
Here $\zeta (\beta ,\ s)$ is the Ruelle zeta function \cite{g},
\begin{equation}
\zeta ^{-1}(\beta ,\ s)=\mathop{\displaystyle\prod}_{p}
\left[ 1-\exp (\beta w_{p}-sn_{p})|\Lambda _{p}|^{-1}\right] ,
\label{Rz}
\end{equation}
where the product is over all the primitive POs, $n_{p}$ is the period of PO
$p$, and $\Lambda _{p}$ is the associated Lyapunov eigenvalue ($|\Lambda
_{p}|>1$). One can express (\ref{Rz}) as a power series in $\exp (-s)$:
\begin{equation}
\zeta ^{-1}(\beta ,\ s)=1+\sum_{n=1}^{\infty }c_{n}(\beta )\exp (-sn),
\label{Rzs}
\end{equation}
where the $n>1$ terms are known as ``curvatures'' \cite{c}. The
convergence of (\ref{Rzs}) is generally better than that of (\ref{Rz}).
Formula (\ref{De}) with a (truncated) curvature expansion (\ref{Rzs}) has
been applied to several systems \cite{a,acl,cgs,e,mr,c,as}. For uniformly
hyperbolic maps with a complete symbolic dynamics (i.e., whose grammar is
unrestricted by ``pruning rules''), all the curvature terms in
(\ref{Rzs}) vanish identically and the exact value of $D$ from
(\ref{De}) coincides precisely with the diffusion rate (\ref{Dn}) for
$n=1$. Examples of such trivial systems are 1D piecewise linear maps
\cite{a,acl} and chains of coupled baker maps \cite{cgs,g}. The
application of (\ref{De}) and other PO formulas \cite{wnv} to a more
realistic system, the periodic Lorentz gas, gives results \cite{mr}
that are within $8\%$ of the values of $D$ obtained by standard methods.
For standard maps (\ref{map}), the quasilinear (strong-chaos) limit of
$D$ is approximated by using just POs of period $n=1$ and $n=2$ \cite{e}.
For the cat and sawtooth maps, the quasilinear approximation of $D$ is
reproduced by formulas related to (\ref {De}) \cite{as}.\newline

In this paper, chaotic diffusion on POs is studied for a nontrivial
Hamiltonian system exhibiting a transition from uniform to nonuniform
hyperbolicity as a parameter is varied. This is the perturbed Arnol'd cat
map on the cylinder, defined by (\ref{map}) with
\begin{equation}
f(x)=f_{0}(x)+\frac{\kappa }{2\pi }\sin (2\pi x),  \label{fx}
\end{equation}
where $f_{0}(x)=x$ for $|x|<0.5$, $f_{0}(-0.5)\equiv 0$,
$f_{0}(x+1)=f_{0}(x) $, and $\kappa $ is a perturbation parameter. This
system, with the definition $f_{0}(-0.5)\equiv -0.5$, is usually
considered on a torus, $-0.5\leq x,\ l<0.5$. Perturbed cat maps on the
torus have attracted much attention recently in the context of ``quantum
chaos'' \cite{pc,pc1,d3}. Anosov theorem \cite{va} states that the
dynamics on the torus for sufficiently small $\kappa$, $\kappa <\kappa
_{{\rm c}}$, is topologically equivalent to that of the unperturbed
($\kappa =0$) system (in particular, the system is completely chaotic
for $\kappa <\kappa _{{\rm c }}$). This expresses the well known {\em
structural stability} of cat maps (see more details in Sec. II, where we
determine $\kappa _{{\rm c}}\approx 0.437$). Actually, we provide
numerical evidence in Sec. II that the structural-stability regime
extends, at least approximately, beyond $\kappa _{{\rm c}}$, up to
$\kappa \approx 1$. A fully chaotic regime is observed up to $\kappa
\approx 1.5$. For larger perturbations, stability islands born by
bifurcation, leading to a significant mixed phase space for
$\kappa > 1.7$ (see Fig. 1). We emphasize that unperturbed cat maps
already feature a very nontrivial symbolic dynamics with nonexplicit
pruning rules given by an infinite set of inequalities \cite{pv1}. As a
result, all the curvature terms in zeta-function expansions are
nonvanishing. See, e.g., an exact expression of $\zeta ^{-1}(0,\ s)$ for
cat maps derived in Sec. II. The relevant dynamics on the cylinder can
be easily inferred from that on the torus.\newline

In Sec. III, we calculate accurately the diffusion coefficient $D$ for
$\kappa $ up to $\kappa \approx 1$ using several PO formulas: (a) The
curvature-expansion formula [(\ref{De}) with (\ref{Rzs})]. (b) The
average of $w_{p}^{2}$ weighted by a stability factor. This formula was
used in Refs. \cite{wnv,mr} and we give a derivation of it from
(\ref{De}). (c) The nonweighted-average formula (\ref{Dn}). The
convergence of each formula as the order $n$ of approximation is
increased is verified. When compared with results obtained by standard
methods, the results from formulas (a) and (b) appear to be more
accurate than those in previous works. The relative difference between
the PO and standard results is not larger than $0.4\%$ when formula (b)
is used and not larger than $1.7\%$ when formula (a) is used. Formula
(c) gives reasonably accurate results (within $2\%$ of the standard
results) for sufficiently small values of $\kappa$ corresponding also to
cases of a considerably nonuniform hyperbolicity. This is due to {\em
uniformity sum rules} satisfied by the PO Lyapunov eigenvalues at {\em
fixed winding number} $w$. These sum rules follow from general arguments
and are supported by much numerical evidence.\newline

\begin{center}
{\bf II. CAT MAPS AND STRUCTURAL STABILITY}\\[0pt]
\end{center}

Consider the unit torus ${\Bbb T}^{2}:-0.5\leq x,\ l<0.5$ and let ${\bf
z} \equiv (x,\ l)$. Hyperbolic cat maps on ${\Bbb T}^{2}$ are defined by
the map $\phi _{0}$: ${\bf z}^{\prime }=A\cdot {\bf z}$ $\mathop{\rm
mod}$ ${\Bbb T}^{2}$, where $A$ is a $2\times 2$ integer matrix with
$\det (A)=1$ and Tr$(A)>2$. While these maps are uniformly hyperbolic,
they feature a very nontrivial symbolic dynamics \cite{pv1}. As a result,
all the curvature terms in zeta-function expansions are nonvanishing. We
show this here by deriving an exact expression of $\zeta ^{-1}(0,\ s)$
for $\phi _{0}$. First, the uniform hyperbolicity implies that the
Lyapunov eigenvalue $\Lambda _{p}$ of any PO with period $n$ is given by
$\Lambda _{p}=\Lambda ^{n}$, where $\Lambda $ is the largest eigenvalue
of $A$. Then, from (\ref {Rz}),
\begin{equation}
\zeta ^{-1}(0,\ s)=\prod_{n=1}^{\infty }\left[ 1-\rho ^{n}(s)\right] ^{N(n)},
\label{z0s}
\end{equation}
where $\rho (s)=\Lambda ^{-1}\exp (-s)$ and $N(n)$ is the number of
primitive POs of period $n$. For $\rho (s)<1$, we find from Rel. (\ref{z0s})
that
\begin{eqnarray}
\ln \left[ \zeta ^{-1}(0,\ s)\right] &=&\sum_{j=1}^{\infty }N(j)\ln \left[
1-\rho ^{j}(s)\right] =  \nonumber \\
-\sum_{j=1}^{\infty }jN(j)\sum_{i=1}^{\infty }\frac{\rho ^{ij}(s)}{ij}
&=&-\sum_{n=1}^{\infty }\frac{\rho ^{n}(s)}{n}\sum_{j\mid n}jN(j),
\label{lz0s}
\end{eqnarray}
where $j\mid n$ means that the positive integer $j$ divides $n$. We now use
the general relation
\begin{equation}
\sum_{j\mid n}jN(j)=P(n),  \label{NPn}
\end{equation}
where $P(n)$ is the number of periodic points of period $n$. For the cat
maps one has \cite{k}
\begin{equation}
P(n)=|\text{Tr}(A^{n})-2|=\Lambda ^{n}+\Lambda ^{-n}-2.  \label{Pn}
\end{equation}
Using (\ref{NPn}) and (\ref{Pn}) in (\ref{lz0s}), we obtain, for $s\geq 0$, 
\begin{equation}
\zeta ^{-1}(0,\ s)=\exp \left[ -\sum_{n=1}^{\infty }P(n)\frac{\rho ^{n}(s)}
{n}\right] =\frac{\left[ 1-\exp (-s)\right] \left[ 1-\Lambda ^{-2}\exp (-s)
\right] }{\left[ 1-\Lambda ^{-1}\exp (-s)\right] ^{2}}.  \label{z0se}
\end{equation}
The explicit curvature expansion of $\zeta ^{-1}(0,\ s)$ for $s\geq 0$ is
easily found from (\ref{z0se}): 
\begin{equation}
\zeta ^{-1}(0,\ s)=1-P(1)\sum_{n=1}^{\infty }n\Lambda ^{-n}\exp (-sn) .
\label{z0sc}
\end{equation}
Thus, all the curvatures $c_{n}(0)$, $n>1$, are nonvanishing.\newline

A perturbed cat map on ${\Bbb T}^{2}$ is given by $\phi _{\kappa }$:
${\bf z}^{\prime }=A\cdot {\bf z}+\kappa {\bf F}({\bf z})$ $\mathop{\rm
mod}$ ${\Bbb T}^{2}$, where ${\bf F}({\bf z})$ is a smooth vector field
periodic on ${\Bbb T}^{2}$. Anosov theorem \cite{va} states that for
sufficiently small $\kappa$, $\kappa <\kappa _{{\rm c}}$, $\phi _{\kappa
}$ is topologically conjugate to $\phi _{0}$ by a continuous
near-identity map $H_{\kappa }$, $\phi _{\kappa }=H_{\kappa }\circ \phi
_{0}\circ H_{\kappa }^{-1}$. Thus, any orbit ${\cal O}_{\kappa }$ of
$\phi _{\kappa }$, in particular a PO, can be written as ${\cal
O}_{\kappa }=H_{\kappa }{\cal O} _{0}$, where ${\cal O}_{0}$ is some
orbit of $\phi _{0}$. We emphasize that the relation $\phi _{\kappa
}=H_{\kappa }\circ \phi _{0}\circ H_{\kappa }^{-1}$ does {\em not} imply
that the Lyapunov eigenvalue $\Lambda _{p}$ of a PO ${\cal O}_{\kappa
}=H_{\kappa }{\cal O}_{0}$ is equal to that of ${\cal O}_{0}$ since the
map $H_{\kappa }$ is not differentiable \cite{va}. The bound $\kappa
_{{\rm c}}$ is generally determined by the inequality: $\max_{{\bf
z}\in {\Bbb T}^{2}}\left( |\kappa \partial {\bf F}/\partial {\bf z}\cdot
{\bf z}|/|{\bf z}|\right) <1-\Lambda ^{-1}$, where $|{\bf
z}|=(x^{2}+l^{2})^{1/2}$. Choosing $A=(2,\ 1;\ 1,\ 1)$ (corresponding to
the Arnol'd cat map \cite{va}) and ${\bf F}({\bf z})=(1/2\pi )\sin (2\pi
x)(1,\ 1)$, we see that $\phi _{\kappa }$ is just the map (\ref{map})
with (\ref{fx}) and the definition $f_{0}(-0.5)\equiv -0.5$ [instead of
$f_{0}(-0.5)\equiv 0$]. In this case, $\Lambda =(3+\sqrt{5})/2$ and
$|\partial {\bf F}/\partial {\bf z}\cdot{\bf z}|=\sqrt{2}|x|\cos (2\pi
x)$. From the inequality above we then get $\kappa <\kappa _{{\rm
c}}\approx 0.437$.\newline

The POs of a perturbed cat map for $\kappa <\kappa _{{\rm c}}$ can be
calculated very accurately as follows. First, the POs ${\cal O}_{0}$ of the
cat map are determined exactly using the techniques in Ref. \cite{pv2}. The
perturbed POs ${\cal O}_{\kappa }=H_{\kappa }{\cal O}_{0}$ are then computed
by applying to ${\cal O}_{0}$\ the map $H_{\kappa }$ constructed iteratively
from the nonlinear functional equation satisfied by it, starting from the
solution $H_{\kappa }^{(0)}$ of the ``homological equation'' (see details in
the proof of Anosov theorem \cite{va}). In this way, we have calculated all
the POs ${\cal O}_{\kappa }=H_{\kappa }{\cal O}_{0}$ of the perturbed
Arnol'd cat map with periods $n\leq 14$ with an accuracy of at least
$10^{-10}$; this accuracy was checked by direct iteration of the map. As a
matter of fact, we found no problems in extending these calculations beyond 
$\kappa _{{\rm c}}\approx 0.437$. We have thus verified that all the POs
${\cal O}_{\kappa }=H_{\kappa }{\cal O}_{0}$ with $n\leq 14$ exist at least 
up to $\kappa =0.946$. In order to check to what extent these POs cover the
set of {\em all} POs with $n\leq 14$ for $\kappa _{{\rm c}}<\kappa \leq
0.946$, we have used the $\zeta ^{-1}(0,\ 0)=0$ test \cite{cgs,e,mr,c}: the
curvature expansion (\ref{Rzs}) for $\zeta ^{-1}(0,\ 0)$, restricted to the
set of POs ${\cal O}_{\kappa }=H_{\kappa }{\cal O}_{0}$, was calculated up
to order $n=1,...,\ 14$ for six values of $\kappa $ uniformly distributed in
the interval $0<\kappa \leq 0.946$. The results are shown in Table I. We see
that the behavior of $\zeta ^{-1}(0,\ 0)$ as $n$ increases is basically the
same for both $\kappa >\kappa _{{\rm c}}$ and $\kappa <\kappa _{{\rm c}}$,
and there is a general trend of $|\zeta ^{-1}(0,\ 0)|$ to decrease. This
indicates that the POs ${\cal O}_{\kappa }=H_{\kappa }{\cal O}_{0}$ form, at
least, a large fraction of all the POs for $\kappa _{{\rm c}}<\kappa \leq
0.946$, i.e., almost no bifurcations take place in this interval. Thus, the
structural-stability regime appears to extend, at least approximately, up to
$\kappa =0.946$. On the basis of this observation, only the POs ${\cal O}
_{\kappa }=H_{\kappa }{\cal O}_{0}$ were used in our calculations for 
$0<\kappa \leq 0.946$.\newline

On the cylinder, the system is described by the map (\ref{map}) with (\ref
{fx}) and $f_{0}(-0.5)\equiv 0$. The discontinuity of (\ref{fx}) at $x=-0.5$
can be viewed as an infinite-slope (vertical) segment. Is is then easy to
show that POs with points on $x=-0.5$ must have an infinite value of
$\Lambda _{p}$. Thus, they will not contribute to (\ref{Rz}) and will not be
considered. Clearly, the initial conditions for the relevant POs (having no
point on $x=-0.5$) can be chosen as points of torus POs ${\cal O}_{\kappa
}=H_{\kappa }{\cal O}_{0}$ lying in the domain $|x|,\ |l|<0.5$. The number
$N(n)$ of relevant POs used in our diffusion calculations in Sec. III is
listed in Table II. The winding number $w_{p}$ is calculated from 
(\ref{PPO}). Strictly speaking, there is {\em no} structural stability on 
the cylinder since $w_{p}$ generally changes when $\kappa $ is varied in
any interval, in particular $(0,\ \kappa _{{\rm c}})$.\newline

\begin{center}
{\bf III. RESULTS FOR }$D${\bf \ AND UNIFORMITY SUM RULES}\\[0pt]
\end{center}

We start by deriving a general formula [Eq. (\ref{DWA}) below] giving an
approximation to $D$ in terms of all the primitive POs of period $n$.  
Let us calculate the derivatives $\partial ^{2}\zeta ^{-1}(\beta ,\
s)/\partial \beta ^{2}$ and $\partial \zeta ^{-1}(\beta ,\ s)/\partial s$
in (\ref{De}) for $\beta =0$ and $s>0$ by direct differentiation of
the infinite product (\ref{Rz}), which is convergent and nonvanishing for
$s>0$ \cite{cgs,g}. Because of the inversion symmetry of (\ref{map})
[$f(-x)=-f(x)$], for each PO with characteristics $(w_{p},\ \Lambda
_{p})$ there exists a PO with characteristics $(-w_{p},\ \Lambda _{p})$.
This implies that $\left. \partial \zeta ^{-1}(\beta ,\ s)/\partial \beta
\right| _{\beta =0,\ s>0}=0$. Using this fact and taking the limit of
$s\rightarrow 0^{+}$, we obtain the following expression for $D$:
\begin{equation}
D=\lim_{n\rightarrow \infty }\frac{\sum_{n^{\prime }=1}^{n}g(n^{\prime })
D_{{\rm WA}}(n^{\prime })}{\sum_{n^{\prime }=1}^{n}g(n^{\prime })},
\label{ADWA}
\end{equation}
where
\begin{equation}
g(n)=n\sum_{p\in {\cal U}_{n}}\frac{|\Lambda _{p}|^{-1}}{1-|\Lambda
_{p}|^{-1}},  \label{gn}
\end{equation}
\begin{equation}
D_{{\rm WA}}(n)=\frac{1}{2g(n)}\sum_{p\in {\cal U}_{n}}\frac{|\Lambda
_{p}|^{-1}}{\left( 1-|\Lambda _{p}|^{-1}\right) ^{2}}\ w_{p}^{2}.
\label{DWA}
\end{equation}
The Hannay-Ozorio-de-Almeida uniformity sum rule \cite{g,hoda} implies
that $\lim_{n\rightarrow \infty }g(n)=1$. Then, the limit in (\ref{ADWA})
exists only if $D_{{\rm WA}}(n)$ converges to $D$. For sufficiently large
$n$, $D_{{\rm WA}}(n)$ in (\ref{DWA}) is just the average of
$w_{p}^{2}/(2n)$ ($p\in {\cal U}_{n}$) weighted by the stability factor
$|\Lambda _{p}|^{-1}$. Such approximations to $D$ have been used in
previous works \cite{wnv,mr}.\newline

Formula (\ref{DWA}) reduces to (\ref{Dn}) in the case of uniform
hyperbolicity. Consider, however, an equivalent expression for
(\ref{DWA}):
\begin{equation}
D_{{\rm WA}}(n)=\frac{1}{2nN(n)}\sum_{w}N_{w}(n)S_{w}(n)w^{2},  \label{DWAw}
\end{equation}
where $N(n)$ and $N_{w}(n)$ are defined as in (\ref{Dn}) and
\begin{equation}
S_{w}(n)=\frac{nN(n)}{g(n)N_{w}(n)}\sum_{p\in {\cal U}_{n},\ w_{p}=w}
\frac{|\Lambda _{p}|^{-1}}{\left( 1-|\Lambda _{p}|^{-1}\right) ^{2}}\ .
\label{Swn}
\end{equation}
The quantity (\ref{Swn}) is a natural restriction of (\ref{gn}) to the
subset of POs with given winding number. If we now assume, in analogy to
$\lim_{n\rightarrow \infty}g(n)=1$, the {\em uniformity sum rules} at
{\em fixed winding number} $w$,
\begin{equation}
\lim_{n\rightarrow \infty }\ S_{w}(n)=1,  \label{usrw}
\end{equation}
formula (\ref{DWAw}) may reduce essentially to (\ref{Dn}) {\em also} in
cases of nonuniform hyperbolicity. We found much numerical evidence for
the validity of (\ref{usrw}) in our system. Part of this evidence is
presented below. In general, the origin of (\ref{usrw}) can be understood
as follows. In our notation, one has the approximate relation for $n\gg 1$
(see, e.g., Appendix B in Ref. \cite{dmss}):
\begin{equation}
\frac{1}{\sqrt{4\pi nD}}\exp \left( -\frac{w^{2}}{4Dn}\right) \approx
n\sum_{n_{p}r=n,\ w_{p}r=w}\frac{|\Lambda _{p}^{r}|^{-1}}{\left( 1-|\Lambda
_{p}^{r}|^{-1}\right) ^{2}},  \label{disw}
\end{equation}
where $r$ (an integer) is the repetition index. The left-hand side of
(\ref {disw}) gives the probability distribution for a generic chaotic
ensemble to diffuse a ``distance'' $|w|$ in ``time'' $n$. Now, as
$n\rightarrow \infty$, there should be no essential difference between
such an ensemble and the PO ensemble ${\cal U}_{n}$. The probability
distribution above is then expected to be approximately equal to
$N_{w}(n)/N(n)$ provided $|w|$ is not too close to the maximal value of
$|w_{p}|$ \cite{d1}. We use this in (\ref{disw}), keeping only the
dominant terms ($r=1$) on the right-hand side. Recalling also the
definition (\ref{Swn}) and the uniformity sum rule $\lim_{n\rightarrow
\infty}g(n)=1$, Rel. (\ref{usrw}) is obtained.\newline

In our numerical calculations we have used the relevant POs on the
cylinder with periods $n\leq 14$, computed as described in Sec. II. The
following PO quantities were calculated accurately for $\kappa =0.086k$,
$k=1,...,\ 11$: (a) $S_{w}(n)$ for all possible values of $w\geq 0$ [note
that $S_{-w}(n)=S_{w}(n)$ from inversion symmetry]. (b) The
curvature-expansion (CE) approximation to $D$, $D_{{\rm CE}}(n)$,
obtained by using in (\ref{De}) the expansion (\ref{Rzs}) truncated after
the first $n$ terms. (c) $D_{{\rm WA}}(n)$ [weighted-average formula
(\ref{DWA})]. (d) $D(n)$ [nonweighted-average formula (\ref{Dn})]. The PO
results for $D$ were compared with standard ones obtained from (\ref{D})
by choosing ${\cal E}$ as the entire unit torus. For this ensemble, which
is obviously invariant under the torus map, one has the exact expansion
\cite{cm}
\begin{equation}
D_{{\cal E}}(m)=\frac{1}{2}C_{0}+\sum_{j=1}^{m-1}\left( 1-\frac{j}{m}\right)
C_{j},\ \ \ \ \ m>1,  \label{DC}
\end{equation}
where $C_{j}=\langle f(x_{0})f(x_{j})\rangle _{{\cal E}}$ are the
force-force correlations for (\ref{fx}). These correlations were calculated
very accurately for $j\leq 30$ by a sophisticated integration of
$f(x_{0})f(x_{j})$ over the unit torus. In general, we found that
$D_{{\cal E}}(m)$ in (\ref{DC}) converges rapidly to $D$ due to the fast 
decay of $C_{j}$. For example, $D_{{\cal E}}(20)$ differs from both 
$D_{{\cal E}}(10)$ and $D_{{\cal E}}(30)$ by no more than $0.05\%$ for all 
the values of $\kappa$ considered. In what follows, $D_{{\cal E}}(20)$
will serve as our ``standard'' value $D_{{\rm S}}$ for $D$.\newline

In Tables III-V we list the four quantities above for three representative
values of $\kappa$; the corresponding value of $D_{{\rm S}}$ is also given.
Table VI shows the quantities $S_{0}(14)$, $D_{{\rm CE}}(14)$, $D_{{\rm WA}
}(14)$, $D(14)$, and $D_{{\rm S}}$ for all the $11$ values of $\kappa $
considered. The results in Tables III-V, as well as similar results for the
other values of $\kappa$, indicate that $D_{{\rm CE}}(n)$, $D_{{\rm WA}}(n)$,
and $D(n)$ start to converge, in general, for $n>8$. These quantities
vanish for $n\leq 3$ since all the POs with period $n\leq 3$ have $w_{p}=0$,
even for $\kappa =0.946$. Table VI shows a very good agreement between $D_{
{\rm WA}}(14)$ and $D_{{\rm S}}$ for all values of $\kappa $. This agreement
is generally better than that between $D_{{\rm CE}}(14)$ and $D_{{\rm S}}$.
The relative difference between $D_{{\rm WA}}(14)$ and $D_{{\rm S}}$ ranges
from $0.04\%$ to $0.4\%$ while that between $D_{{\rm CE}}(14)$ and $D_{{\rm S
}}$ ranges from $0.16\%$ to $1.7\%$.\newline

Table VI also shows that the relative difference between the values of
$D(14)$ and $D_{{\rm S}}$ for $\kappa \leq 0.43$ is not larger than
$2\%$ despite the fact that the hyperbolicity for $\kappa \leq 0.43$ can
be considerably nonuniform, see Fig. 2. To understand this, consider the
behavior of $S_{w}(n)$ for $\kappa \leq 0.43$ in Tables III and IV. While
the convergence of $S_{w}(n)$ to $1$ for $|w|\leq 2$ is quite evident,
$S_{3}(14)$ and $S_{4}(14)$ are not sufficiently converged due to the
relatively small number of POs with $w=3,\ 4$. Precisely because of this
last fact, however, the effect of $S_{\pm 3}(14)-1$ and $S_{\pm 4}(14)-1$
in (\ref{DWAw}) is not significant, leading to only a small difference
between $D(14)$ and $D_{{\rm WA}}(14)$ (or $D_{{\rm S}}$). As $\kappa $
is increased, a larger value of $|w|$ ($|w|=5$) appears for $\kappa
>0.516$ and the hyperbolicity becomes more nonuniform. Then, since the
total number $N(n)$ of POs is constant (does not depend on $\kappa$), the
convergence of $S_{w}(n)$ is expected to deteriorate for all $w$. This
is, in fact, confirmed by all our numerical data (see the worse case for
$\kappa =0.946$ in Table V) with the exception of $w=0$. As Table VI
shows, $S_{0}(14)$ remains remarkably well converged for all values of
$\kappa$. We shall attempt to find an explanation of this and other
facts in a future work.\newline

\begin{center}
{\bf IV. CONCLUSIONS}\\[0pt]
\end{center}

In this paper, chaotic diffusion on periodic orbits (POs) was studied for
the perturbed Arnol'd cat map on the cylinder, in a relatively large
range of perturbations $\kappa $ corresponding to a
``structural-stability'' regime of the system on the torus. Numerical
evidence indicates that this regime extends, at least approximately,
significantly beyond the Anosov bound $\kappa _{{\rm c}}\approx 0.437$,
i.e., at least up to $\kappa \approx 1$. This extension, which was
already noticed in a quantum-chaos context for a different perturbed cat
map \cite{pc}, is further supported by the very good agreement between
the PO and standard results for $D$ also for $\kappa _{{\rm c}}<\kappa
\leq 0.946$, where only POs topologically conjugate to the $\kappa =0$
POs are used. In the absence of bifurcations, the variation of $D$ with
$\kappa $ is totally due to the change of the characteristics $(w_{p},\
\Lambda _{p})$ of a constant number of POs. Thus, the case studied in
this paper is basically different from that considered in Ref. \cite{e},
i.e., standard maps in a strong-chaos limit. In the latter case,
bifurcations of low-period ($n=1$ and $n=2$) POs are the main cause for
the relevant variation of $D$ with the parameter. The usual distinction
between fundamental ($n=1$) and curvature ($n>1$) terms \cite{c} is not
felt in our case since $w_{p}=0$ for $n\leq 3$ and for all the values of
$\kappa $ considered. The convergence of the PO results to $D$ generally
starts only for order (or period) $n>8$.\newline

As in the case of the Lorentz gas \cite{mr}, the most accurate PO results
are obtained by using the weighted-average formula (\ref{DWA}). In
general, by expressing this formula in the winding-number representation
(\ref{DWAw}), it becomes clear that the effect of a nonuniform
hyperbolicity is completely captured by the basic quantities (\ref{Swn}).
Since these quantities satisfy the uniformity sum rules (\ref{usrw}), the
manifestation of this effect is essentially restricted, for sufficiently
large $n$, to the ``tail'' of the distribution (\ref{disw}) ($|w|$ large
relatively to the maximal value of $|w_{p}|$). Here the discrepancy
between $N_{w}(n)/N(n)$ and (\ref{disw}) leads to a value of $S_{w}(n)$
which is not well converged. We have shown that the effect of
nonconverged values of $S_{w}(n)$ may be insignificant also in cases of a
considerably nonuniform hyperbolicity. Then, formula (\ref{DWA}) reduces
essentially to the nonweighted-average formula (\ref{Dn}).\newline

{\bf Acknowledgments}\newline

We thank J.M. Robbins for discussions. This work was partially supported
by the Israel Science Foundation administered by the Israel Academy of
Sciences and Humanities. V.E.C. acknowledges the CRDF and Ministry of
Education of the Russian Federation for Award \#VZ-010-0.

\newpage

Table I. $\zeta ^{-1}(0,\ 0)$ to order $n=1,...,\ 14$ of curvature expansion
for several values of $\kappa $.
\[
\begin{tabular}{|r|r|r|r|r|r|r|}
\hline
$n$ & \multicolumn{6}{|c|}{$\zeta ^{-1}(0,\ 0)$} \\ \cline{2-7}
& $\kappa =0.086$ & $\kappa =0.258$ & $\kappa =0.43$ & $\kappa =0.602$ &
$\kappa =0.774$ & $\kappa =0.946$ \\ \hline
$1$ & $-0.36790$ & $-0.34306$ & $-0.32172$ & $-0.30313$ & $-0.28676$ &
$-0.27220$ \\
$2$ & $-0.29773$ & $-0.31199$ & $-0.33060$ & $-0.35530$ & $-0.38853$ &
$-0.43399$ \\
$3$ & $-0.17097$ & $-0.17537$ & $-0.17565$ & $-0.17293$ & $-0.16915$ &
$-0.16734$ \\
$4$ & $-0.08734$ & $-0.09026$ & $-0.09120$ & $-0.08990$ & $-0.08641$ &
$-0.08143$ \\
$5$ & $-0.04179$ & $-0.04339$ & $-0.04390$ & $-0.04242$ & $-0.03686$ &
$-0.02308$ \\
$6$ & $-0.01918$ & $-0.02003$ & $-0.02043$ & $-0.01995$ & $-0.01786$ &
$-0.01328$ \\
$7$ & $-0.00856$ & $-0.00899$ & $-0.00925$ & $-0.00909$ & $-0.00795$ &
$-0.00478$ \\
$8$ & $-0.00374$ & $-0.00395$ & $-0.00410$ & $-0.00408$ & $-0.00360$ &
$-0.00222$ \\
$9$ & $-0.00342$ & $-0.00688$ & $-0.00995$ & $-0.01261$ & $-0.01476$ &
$-0.01621$ \\
$10$ & $-0.00002$ & $0.00104$ & $0.00185$ & $0.00249$ & $0.00306$ & $0.00375$
\\
$11$ & $0.00025$ & $0.00130$ & $0.00237$ & $0.00350$ & $0.00480$ & $0.00643$
\\
$12$ & $-0.00002$ & $0.00020$ & $0.00042$ & $0.00059$ & $0.00073$ & $0.00087$
\\
$13$ & $0.00019$ & $0.00061$ & $0.00097$ & $0.00125$ & $0.00147$ & $0.00164$
\\
$14$ & $0.00012$ & $0.00038$ & $0.00062$ & $0.00083$ & $0.00099$ & $0.00103$
\\ \hline
\end{tabular}
\]

\[
\]

Table II. Number $N(n)$ of relevant POs of period $n$, having no point on 
$x=-0.5$.
\[
\begin{tabular}{|c|c|c|c|c|c|c|c|c|c|c|c|c|c|c|}
\hline
$n$ & $1$ & $2$ & $3$ & $4$ & $5$ & $6$ & $7$ & $8$ & $9$ & $10$ & $11$ &
$12$ & $13$ & $14$ \\ \hline
$N(n)$ & $1$ & $2$ & $2$ & $10$ & $24$ & $48$ & $120$ & $270$ & $568$ &
$1500$ & $3600$ & $8543$ & $20880$ & $50700$ \\ \hline
\end{tabular}
\]

\newpage

Table III. $S_{w}(n)$, $D_{{\rm CE}}(n)$, $D_{{\rm WA}}(n)$, and $D(n)$ for
$\kappa =0.258$; $D_{{\rm S}}=0.04865$.
\[
\begin{tabular}{|r|r|r|r|r|r|r|r|r|}
\hline
$n$ & $S_{0}(n)$ & $S_{1}(n)$ & $S_{2}(n)$ & $S_{3}(n)$ & $S_{4}(n)$ & $D_{
{\rm CE}}(n)$ & $D_{{\rm WA}}(n)$ & $D(n)$ \\ \hline
1 & $1.52221$ &  &  &  &  & $0.00000$ & $0.00000$ & $0.00000$ \\
2 & $1.19039$ &  &  &  &  & $0.00000$ & $0.00000$ & $0.00000$ \\
3 & $1.05180$ &  &  &  &  & $0.00000$ & $0.00000$ & $0.00000$ \\
4 & $1.01987$ & $1.02505$ &  &  &  & $0.02724$ & $0.05125$ & $0.05000$ \\
5 & $1.03504$ & $0.97134$ &  &  &  & $0.03261$ & $0.04047$ & $0.04167$ \\
6 & $1.02088$ & $0.98225$ &  &  &  & $0.03350$ & $0.03752$ & $0.03819$ \\
7 & $1.01632$ & $0.96115$ & $1.13505$ &  &  & $0.04102$ & $0.05023$ & 
$0.04881$ \\
8 & $1.02287$ & $0.96467$ & $1.06022$ &  &  & $0.04410$ & $0.04795$ & 
$0.04768$ \\
9 & $0.99896$ & $0.99316$ & $1.04018$ &  &  & $0.04487$ & $0.04762$ &
$0.04695$ \\
10 & $1.01179$ & $0.98401$ & $1.01905$ & $1.24670$ &  & $0.04648$ & $0.04807$
& $0.04760$ \\
11 & $1.00966$ & $0.98787$ & $1.00285$ & $1.17419$ &  & $0.04726$ & $0.04862$
& $0.04811$ \\
12 & $1.00935$ & $0.98897$ & $1.00102$ & $1.11421$ &  & $0.04751$ & $0.04819$
& $0.04776$ \\
13 & $1.00981$ & $0.98864$ & $0.99965$ & $1.08830$ & $1.40865$ & $0.04801$ &
$0.04873$ & $0.04826$ \\
14 & $1.00658$ & $0.99180$ & $0.99950$ & $1.05561$ & $1.30305$ & $0.04824$ &
$0.04861$ & $0.04815$ \\ \hline
\end{tabular}
\]

\newpage

Table IV. $S_{w}(n)$, $D_{{\rm CE}}(n)$, $D_{{\rm WA}}(n)$, and $D(n)$ for
$\kappa =0.43$; $D_{{\rm S}}=0.05380$.
\[
\begin{tabular}{|r|r|r|r|r|r|r|r|r|}
\hline
$n$ & $S_{0}(n)$ & $S_{1}(n)$ & $S_{2}(n)$ & $S_{3}(n)$ & $S_{4}(n)$ & $D_{
{\rm CE}}(n)$ & $D_{{\rm WA}}(n)$ & $D(n)$ \\ \hline
1 & $1.47432$ &  &  &  &  & $0.00000$ & $0.00000$ & $0.00000$ \\
2 & $1.21592$ &  &  &  &  & $0.00000$ & $0.00000$ & $0.00000$ \\
3 & $1.04705$ &  &  &  &  & $0.00000$ & $0.00000$ & $0.00000$ \\
4 & $1.03876$ & $0.99752$ &  &  &  & $0.02630$ & $0.04988$ & $0.05000$ \\
5 & $1.06682$ & $0.92808$ &  &  &  & $0.03128$ & $0.03867$ & $0.04167$ \\
6 & $1.02173$ & $0.95875$ & $1.27251$ &  &  & $0.03923$ & $0.05429$ &
$0.05208$ \\
7 & $1.02423$ & $0.95049$ & $1.13318$ &  &  & $0.04692$ & $0.05640$ &
$0.05476$ \\
8 & $1.02965$ & $0.95429$ & $1.07500$ &  &  & $0.04871$ & $0.05239$ &
$0.05185$ \\
9 & $0.99527$ & $0.98363$ & $1.06576$ & $1.74814$ &  & $0.05076$ & $0.05433$
& $0.05184$ \\
10 & $1.02227$ & $0.97717$ & $0.98799$ & $1.35259$ &  & $0.05216$ & $0.05339$
& $0.05253$ \\
11 & $1.01290$ & $0.98137$ & $1.00161$ & $1.23369$ &  & $0.05265$ & $0.05426$
& $0.05308$ \\
12 & $1.01111$ & $0.98476$ & $1.00131$ & $1.13098$ & $2.06872$ & $0.05294$ &
$0.05318$ & $0.05216$ \\
13 & $1.00872$ & $0.98746$ & $0.99941$ & $1.08594$ & $1.59230$ & $0.05323$ &
$0.05372$ & $0.05271$ \\
14 & $1.00869$ & $0.98753$ & $0.99789$ & $1.07197$ & $1.43866$ & $0.05338$ &
$0.05378$ & $0.05273$ \\ \hline
\end{tabular}
\]

\newpage

Table V. $S_{w}(n)$, $D_{{\rm CE}}(n)$, $D_{{\rm WA}}(n)$, and $D(n)$ for
$\kappa =0.946$; $D_{{\rm S}}=0.07200$.
\[
\begin{tabular}{|r|r|r|r|r|r|r|r|r|r|}
\hline
$n$ & $S_{0}(n)$ & $S_{1}(n)$ & $S_{2}(n)$ & $S_{3}(n)$ & $S_{4}(n)$ &
$S_{5}(n)$ & $D_{{\rm CE}}(n)$ & $D_{{\rm WA}}(n)$ & $D(n)$ \\ \hline
1 & $1.37400$ &  &  &  &  &  & $0.00000$ & $0.00000$ & $0.00000$ \\
2 & $1.42345$ &  &  &  &  &  & $0.00000$ & $0.00000$ & $0.00000$ \\
3 & $1.03401$ &  &  &  &  &  & $0.00000$ & $0.00000$ & $0.00000$ \\
4 & $1.19364$ & $0.77294$ &  &  &  &  & $0.01918$ & $0.03865$ & $0.05000$ \\
5 & $1.05191$ & $0.75048$ & $2.07839$ &  &  &  & $0.05413$ & $0.10055$ &
$0.07500$ \\
6 & $1.14789$ & $0.70455$ & $1.63690$ &  &  &  & $0.06132$ & $0.06993$ &
$0.06250$ \\
7 & $1.06410$ & $0.72943$ & $1.60419$ &  &  &  & $0.06600$ & $0.08195$ &
$0.06667$ \\
8 & $1.12569$ & $0.80405$ & $1.16902$ & $1.83040$ &  &  & $0.06826$ &
$0.06781$ & $0.06389$ \\
9 & $0.94494$ & $0.90200$ & $1.33077$ & $1.93640$ &  &  & $0.07233$ &
$0.08259$ & $0.06494$ \\
10 & $1.00057$ & $0.98694$ & $0.91645$ & $1.64065$ &  &  & $0.07232$ &
$0.06972$ & $0.06433$ \\
11 & $0.96412$ & $0.96616$ & $1.09809$ & $1.32642$ & $1.94196$ &  & $0.07138$
& $0.07231$ & $0.06485$ \\
12 & $1.01872$ & $0.94809$ & $1.02208$ & $1.19621$ & $2.92740$ &  & $0.07175$
& $0.07120$ & $0.06448$ \\
13 & $0.97018$ & $0.95294$ & $1.10901$ & $1.09196$ & $2.04263$ & $3.55700$ &
$0.07181$ & $0.07262$ & $0.06507$ \\
14 & $1.00396$ & $0.94759$ & $1.04419$ & $1.11549$ & $1.90135$ & $2.53191$ &
$0.07188$ & $0.07180$ & $0.06505$ \\ \hline
\end{tabular}
\]

\newpage

Table VI. $S_{0}(14)$, $D_{{\rm CE}}(14)$, $D_{{\rm WA}}(14)$, $D(14)$, and
$D_{{\rm S}}$ for all the $11$ values of $\kappa $.
\[
\begin{tabular}{|l|l|l|l|l|l|}
\hline
$\kappa $ & $S_{0}(14)$ & $D_{{\rm CE}}(14)$ & $D_{{\rm WA}}(14)$ & $D(14)$
& $D_{{\rm S}}$ \\ \hline
$0.086$ & $1.00193$ & $0.04350$ & $0.04377$ & $0.04356$ & $0.04388$ \\
$0.172$ & $1.00431$ & $0.04551$ & $0.04618$ & $0.04579$ & $0.04622$ \\
$0.258$ & $1.00658$ & $0.04824$ & $0.04861$ & $0.04815$ & $0.04865$ \\
$0.344$ & $1.00752$ & $0.05070$ & $0.05109$ & $0.05046$ & $0.05116$ \\
$0.43$ & $1.00869$ & $0.05338$ & $0.05378$ & $0.05273$ & $0.05380$ \\
$0.516$ & $1.00807$ & $0.05589$ & $0.05642$ & $0.05506$ & $0.05656$ \\
$0.602$ & $1.00058$ & $0.05841$ & $0.05938$ & $0.05736$ & $0.05942$ \\
$0.688$ & $0.99577$ & $0.06220$ & $0.06228$ & $0.05921$ & $0.06240$ \\
$0.774$ & $1.00191$ & $0.06531$ & $0.06548$ & $0.06102$ & $0.06551$ \\
$0.86$ & $1.00014$ & $0.06838$ & $0.06848$ & $0.06297$ & $0.06873$ \\
$0.946$ & $1.00396$ & $0.07188$ & $0.07180$ & $0.06505$ & $0.07200$ \\ \hline
\end{tabular}
\]

\figure{Fig. 1. Mixed phase space for the perturbed Arnol'd cat map
[(\ref{map}) with (\ref{fx})] for $\kappa =2$.}

\figure{Fig. 2. Distributions of the Lyapunov exponents $\lambda_p=\ln
(|\Lambda_p|^{-1})/n$ for all the primitive POs of period $n=14$ of the
perturbed Arnol'd cat map for $\kappa =0.258$ (filled circles), $\kappa
=0.43$ (crosses), and $\kappa =0.946 $ (triangles). Each distribution was
calculated by dividing the full range of $\lambda_p$ into $100$ equal
intervals and counting the number $\Delta N$ of values of $\lambda_p$ in
each interval.}

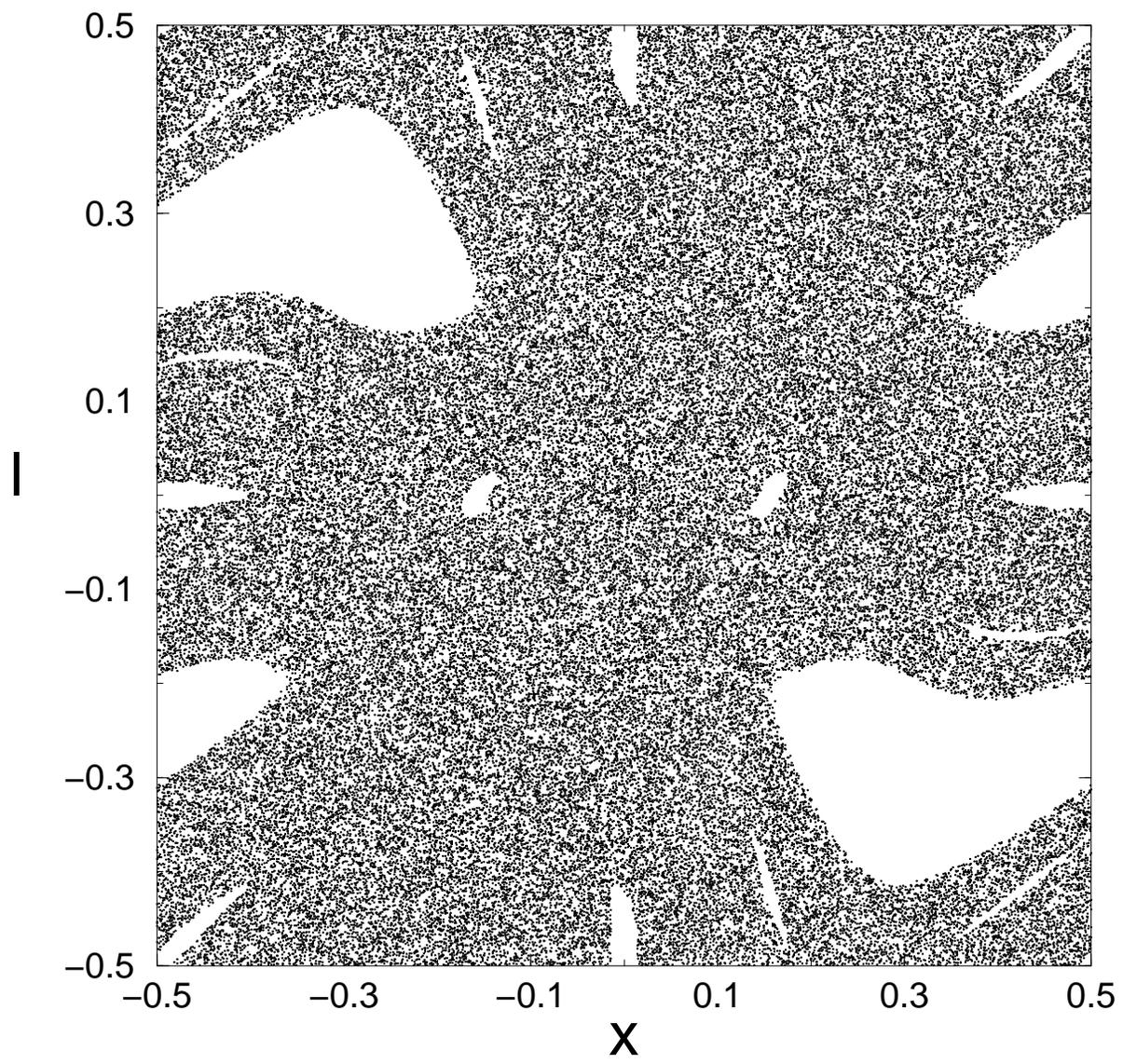
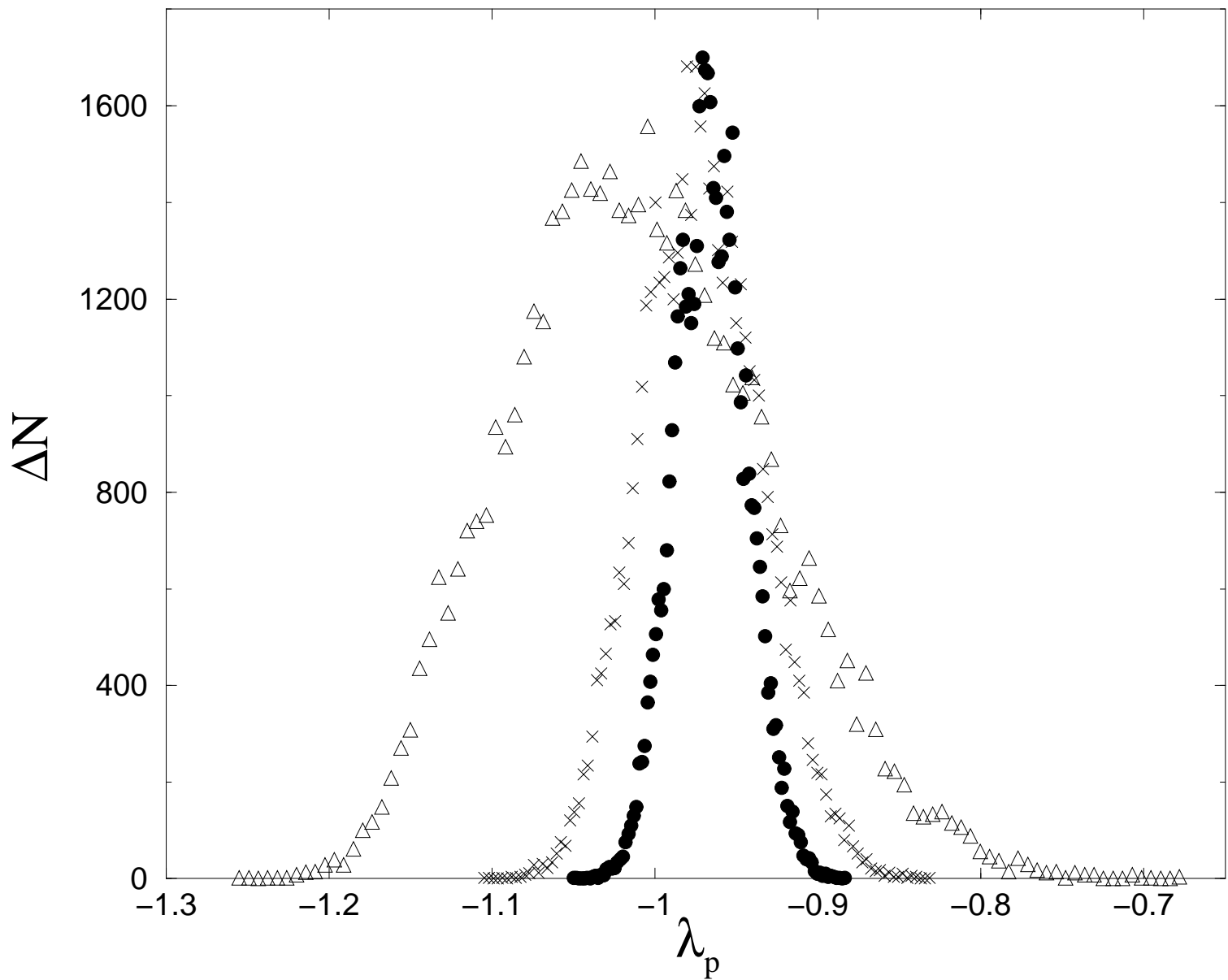
\end{document}